# Intra-connected three-dimensionally isotropic bulk negative index photonic metamaterial

**Durdu Ö. Güney,**[1,*] **Thomas Koschny,**[1,2] **and Costas M. Soukoulis**[1,2]

[1]*Ames National Laboratory, USDOE and Department of Physics and Astronomy, Iowa State University, Ames, IA 50011*

[2] *Institute of Electronic Structure and Laser, Foundation for Research and Technology Hellas (FORTH), and Department of Materials Science and Technology, University of Crete, 7110 Heraklion, Crete, Greece*

[*]dguney@ameslab.gov

**Abstract:** Isotropic negative index metamaterials (NIMs) are highly desired, particularly for the realization of ultra-high resolution lenses. However, existing isotropic NIMs function only two-dimensionally and cannot be miniaturized beyond microwaves. Direct laser writing processes can be a paradigm shift toward the fabrication of three-dimensionally (3D) isotropic bulk optical metamaterials, but only at the expense of an additional design constraint, namely connectivity. Here, we demonstrate with a proof-of-principle design that the requirement connectivity does not preclude fully isotropic left-handed behavior. This is an important step towards the realization of bulk 3D isotropic NIMs at optical wavelengths.

©2010 Optical Society of America

**OCIS codes:** (160.3918) Metamaterials; (220.0220) Optical design and fabrication.

## 1. Introduction

Negative index metamaterials (NIMs) potentially enable many exotic applications as varied as perfect lens [1], invisibility cloaks [2.3], and quantum levitation [4]. Particularly for ultra-high resolution lenses, isotropy is inevitable. Existing only two-dimensionally (2D), isotropic NIMs cannot be miniaturized beyond microwaves [5]. Direct laser writing (DLW) with chemical vapor deposition appears to be a more promising route toward the fabrication of three-dimensionally (3D) isotropic bulk optical metamaterials [6]. However, DLW brings a new design constraint, namely connectivity. We recently proposed the first connected 2D isotropic photonic NIM with about 20THz bandwidth [7]. It is an open question, whether there is any 3D isotropic metamaterial design working for optical frequencies still feasible to fabricate. Here, we report a fully isotropic photonic NIM and demonstrate as a proof-of-principle that the requirement connectivity does not preclude left-handed behavior. This is an important step towards the realization of bulk 3D isotropic NIMs at optical wavelengths.

Since no isotropic magnetic material exists in nature, which possesses negative permeability at THz frequencies, such single-negative metamaterials would be of interest by itself for imaging, magnetic shielding, and other practical applications. Gay-Balmaz and Martin [8] experimentally demonstrated the first 2D isotropic magnetic metamaterial formed by two crossed split-ring-resonators (SRRs). Recently, Padilla [9] and Baena et al. [10] independently proposed fully isotropic bulk magnetic metamaterial designs, based on SRRs arranged in a cubic lattice by employing the spatial symmetries.

On the other hand, some designs of 3D isotropic NIMs exist, but fabricating them has remained a challenging task and virtually impossible at optical frequencies. Koschny et al. [11] for example, designed an early example of an isotropic NIM. However, high-constant dielectric assumed across the gaps of the SRRs render the experimental realization impractical. Alternative approaches have also been investigated including transmission lines [12,13], chiral resonators [14], and high refractive index spheres [15]; but, no practical feasible designs have yet occurred, operating at frequencies beyond the microwave ranges.

The DLW processes, assisted by chemical vapor deposition (CVD), offer a viable means to fabricate metamaterial structures at THz and optical regions by enabling computer-controlled formation of almost arbitrary three-dimensional patterns. These are not possible to fabricate with traditional photolithographic processes [6]. Structural length scales of generated individual 3D materials can span six orders of magnitude, from tens of nanometers to millimeters [6,16-18]. DLW produces a binary structure defined by the path exposed to the laser focus (which polymerizes and remains standing) within the homogeneous background of unexposed photoresist (which dissolves). After the unexposed photoresist has been removed during development, the remaining structure can be metalized by CVD or entirely replaced by metal. To fabricate large-scale bulk metamaterial structures, interconnected unit cells are desired. This helps avoid the collapse of otherwise disconnected and unsupported metallic unit cell components, left free-standing after the unexposed photoresist has been removed. Moreover, connectivity, in general, provides an advantage for rapid prototyping of arbitrary 3D metallic structures using DLW and CVD.

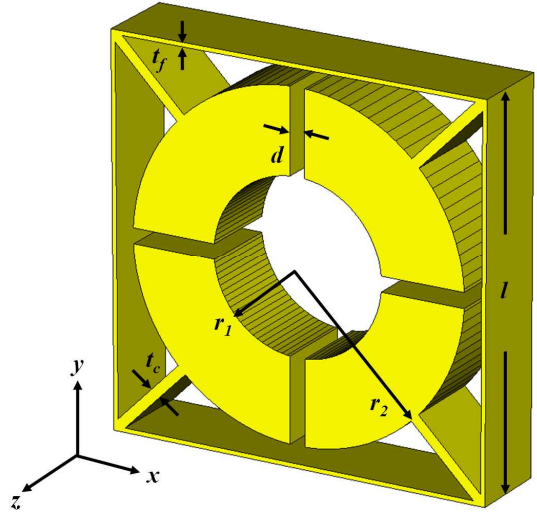

Fig. 1. Basic building block of the 3D isotropic intra-connected metamaterial structure consisting of a four-gap SRR connected diagonally to an outer square frame, all metal (Au) in a vacuum background. $r_1$=65nm (inner ring radius), $r_2$=128nm (outer ring radius), $l$=289nm (frame length), $t_c$=8nm (connector thickness), $t_f$=3nm (frame thickness), and $d$=3nm (gap width).

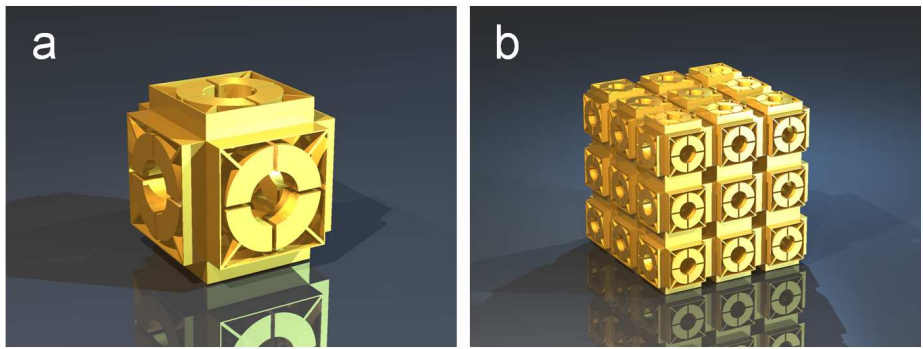

Fig. 2. Unit cell (**a**) and the 3×3×3 bulk (**b**) illustration of the designed intra-connected isotropic NIM structure. Full connectivity is achieved by diagonal connectors and square frames.

From the design perspective, it is a very difficult task to determine a connected structure amenable to fabrication and simultaneously exhibiting acceptable metamaterial properties, particularly at optical frequencies. The magnetic and electric constitutes of the intra-connected metamaterial structure can be thought of as forming an integrated nanocircuitry, which needs a careful design to prevent any unwanted short circuits. Most common NIM geometries, SRRs or fishnet structures [19] with their current design features cannot be assembled into spatially isotropic metamaterials for fabrication at THz frequencies and yet maintain the desired optical properties unless their constitutes are cleverly interconnected. In this letter we attack exactly this important problem and present the first, fully intra-connected and isotropic metamaterial design, principally amenable to fabrication by DLW and CVD. Using rotation symmetries of the cube, our metamaterial allows left-handed behavior for any orthogonal direction of propagation and any polarization of light. Although the size requirements make the designed structure unfeasible to fabricate with the state-of-the-art DLW processes, it provides a proof-of-principle demonstration on how to avoid such hurdles in future isotropic photonic metamaterial designs.

## 2. Three-dimensional isotropic design

In Fig. 1 we show the basic building block of the unit cell. It is a contiguous metallic structure consisting of a bulky four-gap circular SRR with a square cross-section connected diagonally to the square outer frame. Geometric design parameters and their descriptions are given in the caption of Fig. 1. The 4-gap SRR provides a resonant negative Lorentzian effective permeability, while the interconnected frame accounts for a negative Drude-like effective permittivity.

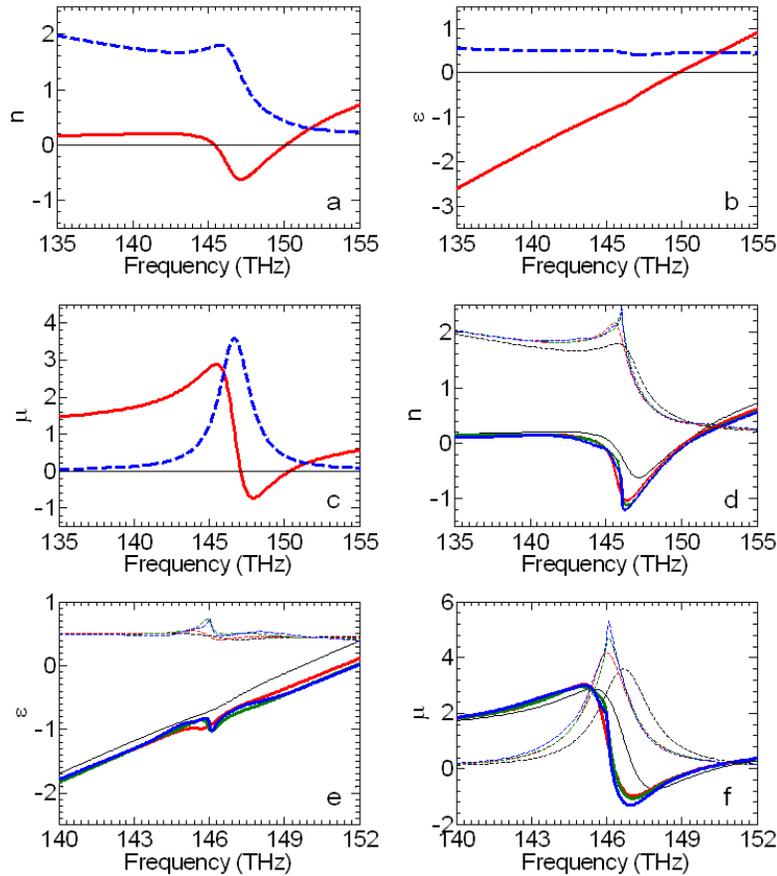

Fig. 3. Retrieved effective parameters $n$, $\varepsilon$, and $\mu$ for a single-unit-cell (**a-c**) and for a different number of unit cells up to four (**d-f**), using the homogeneous effective medium approximation. The solid curves indicate real parts and the dashed curves indicate imaginary parts. In the bottom panels, black, red, green, and blue correspond to 1-4 unit cells, respectively.

We assemble this basic building block to a cubic unit cell illustrated in Fig. 2(a), which, in turn, is arranged into a cubic lattice to form the actual bulk metamaterial in Fig. 2(b). Note, the bulk structure is fully intra-connected and is spatially isotropic for all principal directions of propagation and any polarization of light. It is able to provide 3D isotropic left-handed behavior.

Connectivity without infringing the SRR resonance is based on results we have originally proposed for reducing the losses above THz frequencies [20]. We have shown that bulky structures (i.e., structures with small effective radius of curvatures) lead to significantly reduced geometric skin depth. Hence they confine the resonant current close to the inner edge of the SRR ring and away from the outer surface. Therefore, metallic contacts, necessary for connectivity, must attach at the outer circumference of the SRR ring sufficiently far from sensitive areas, such as the inner rim and the gaps. This leaves us with diagonal connecters as shown in Fig. 1. This is the essence of how to interconnect without destroying the magnetic resonance.

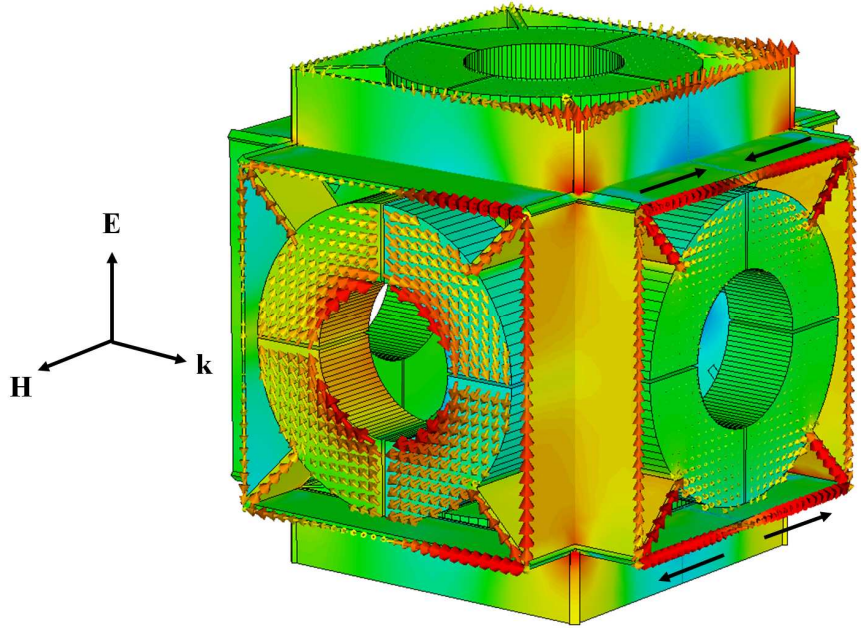

Fig. 4. Total current density distribution around magnetic resonance in logarithmic scale. Size and color of the arrows give the magnitude of the current density at the given position. The background color shows the vertical component of the total current density (orange up, blue down, and green 0). Black arrows are drawn for clarification to improve the readability of the direction of color arrows at the corresponding locations. The Incident field configuration is shown on the left.

To demonstrate the idea, we performed simulations with the CST MICROWAVE STUDIO software package (Computer Simulation Technology GmbH, Darmstadt, Germany). We used the Drude model to describe the metal with the experimental values for the plasma and the collision frequencies for bulk gold [21] ($\omega_p=2\pi\times2.175\times10^{15}$ s$^{-1}$, $\omega_c=2\pi\times6.5\times10^{12}$ s$^{-1}$).

Given their sufficiently subwavelength unit cell sizes, metamaterials can be treated as homogeneous effective media so they can be described by effective electric permittivity ($\varepsilon$), magnetic permeability ($\mu$). Hence, an effective refractive index ($n$) [22]. Figure 3 shows the retrieved effective $n$, $\varepsilon$, and $\mu$ for our structure, which has about a five times smaller unit cell size than the relevant vacuum wavelength. The top panels show the effective parameters for a single-unit-cell only, while the lower panels also include the multiple-unit-cell retrievals up to four unit cells.

Notice the structure homogenizes well after a few unit cells evident from the converging effective parameters. Although for the single unit cell, $n$, does not reach the value of -1, for multiple unit cells (i.e., bulk material), $n$ clearly exceeds -1 and converges, while the corresponding magnetic resonance becomes sharper. The structure visible in $\varepsilon$ is due to periodic effects [23]. For the 4-unit cell structure, $\varepsilon$ and $\mu$ are simultaneously negative over a 4THz bandwidth with a center frequency of 148THz and a figure of merit (FOM) close to unity. The FOM is defined as the ratio $-\Re(n)/\Im(n)$ at $\Re(n)=-1$. Another observation is that multiple unit cell resonance peaks are overall red-shifted from single unit cell resonance due to the coupling between neighboring unit cells. Insignificant variations for systems longer than 2 unit cells can be attributed to numerical error and decreased mesh density for longer structures constrained by computer memory.

In Fig. 4 we show the total current density distribution (log scale) around the magnetic resonance. Because of the cube's symmetry, the current distributions for other orthogonal directions of propagation and polarization are identical up to the geometric global phase. Arrows indicate the total current density at three shown faces of the cube.

The non-resonant negative electric response [see Figs. 3(b) and 3(e)] arises mainly from the parallel currents oscillating along the continuous frame patches in the vertical direction (see orange columns in Fig. 4). On the other hand, resonant magnetic response manifests itself as a circular current (clock-wise), clearly visible at the inner edge of the SRRs at the two faces of the cube perpendicular to the incident **H**-field. There is also an opposing (counter-clock-wise) circular current along the attached square frame. The latter is a diamagnetic response generated by the changing flux of the external **H** and the resonant magnetic response of the SRR, the former due to Lenz's law. Notice on the right side of the frame the electric response current is parallel to the magnetic response current and, therefore, reinforces the net current flow. On the left side, however, they are anti-parallel and therefore we observe a destructive interference.

Other details noticeable in Fig. 4 include (i) charge accumulation on the middle of the frame patches across the SRR gaps, reminiscent of a high order resonance (see blue spots on the top and the bottom of the cube), (ii) electric polarization currents on the connectors, and (iii) high concentrations of current density at the corners (see Ref. 20 for its impact on ohmic losses).

## 3. Conclusion

In summary, the fabrication of a fully isotropic photonic NIM structures is a great challenge. DLW processes with CVD, seen as the most promising technique so far [6], require correct connectivity of the metallic components without short-circuiting the fragile nano-circuits. Here, we have presented for the first time intra-connected 3D isotropic bulk NIM design working around optical wavelengths. The idea of proper connectivity, which is behind our proof-of-principle design, can enable the fabrication of metallic metamaterial structures in the optical region and open a new avenue in the design and fabrication of functional metamaterial devices to allow the unprecedented manipulation of light.

**Acknowledgments**

Work at Ames Laboratory was supported by the Department of Energy (Basic Energy Sciences) under contract No. DE-AC02-07CH11358. This work was partially supported by the European Community FET project PHOME (Contract No. 213390).